\documentclass[aps,preprint,floatfix,altaffilletter]{revtex4}

\usepackage[dvips]{graphicx}
\usepackage{color}
\usepackage{amsmath}
\usepackage{here}

\begin{document}

\begin{sloppy}

\title{Optical isolation with epsilon-near-zero metamaterials}

\author{Arthur R. Davoyan$^*$, Ahmed Mahmoud$^{b}$, Nader Engheta$^{c}$}
\address{Department of Electrical and Systems Engineering, University of Pennsylvania, Philadelphia,
Pennsylvania 19104, USA }
\address{$^*$davoyan@seas.upenn.edu}
\address{$^b$ amahmo$@$seas.upenn.edu,$^c$engheta$@$seas.upenn.edu}
\begin{abstract}
We suggest a principle for isolation of circularly polarized waves in magnetically active extreme-parameter metamaterials. Using theoretical analysis and numerical simulations, we show that metamaterials with extreme parameters, such as epsilon-near-zero materials (ENZ), when merged with magneto-optical materials, become transparent for forward circularly polarized waves of a given handedness and opaque for backward propagating waves of the same handedness. We theoretically study two possible implementations of such hybrid materials: (1) the case of metal-dielectric stacks; and (2) rectangular waveguide near its cut-off frequency. We prove that these structures can be utilized as compact isolators for circularly polarized waves.
\end{abstract}
\maketitle


\section{Introduction}

Many objects in nature ranging from molecules and biological species to astrophysical bodies exhibit different sensitivity to left- and right-handed circular polarizations of light~\cite{Sparks_PNAS,Degtjarev,Kolokolova}. Therefore the ability to filter and isolate circular polarizations is of practical importance for science and engineering. Progress with chiral heterostructures~\cite{Linden,Plum,Helgert,Alu,Giessen}, paved the way for the design and engineering of ultra-compact composite structures distinguishing left- and right-handed polarization states. Nevertheless, any chiral metastructure satisfying the Lorentz reciprocity theorem~\cite{Landau} is reciprocal, implying that interaction of the backward and forward propagation of the same circularly polarized states with such structures are identical and can not be distinguished. Imagine, for example, a right-handed circularly polarized (RCP) light incident normally in $+z$ (forward) direction on a slab of chiral material. Its transmission and reflection amplitudes would differ from left-handed circularly polarized (LCP) light incident from the same direction. At the same time due to reciprocity of the system RCP light propagating backward (in $-z$ direction) experiences exactly the same transmission and reflection as forward propagating RCP, see schematic in Fig.~\ref{slabs}(a). Hence, clearly the isolation and separation of forward and backward waves is not possible in such structures. It is worth mentioning here the works on asymmetric transmission with some types of chiral heterostructures~\cite{Fedotov,Schwanecke,Menzel}, however these structures still satisfy the Lorentz reciprocity conditions and can not be considered as truly nonreciprocal systems with asymmetric transmission.

Difference in forward and backward light flow directions is closely related with the time reversal symmetry breaking of Maxwell's equations~\cite{Haldane,Soljacic} and nonreciprocity of electromagnetic laws~\cite{Landau,Potton}. The latter is possible with either time noninvariant, including gyrotropic, or nonlinear materials. In particular, recent advances with magneto-photonic crystals~\cite{Fan,Khanikaev} and nonlinear dielectric stacks~\cite{Tocci,Miroshnichenko} demonstrate the possibility for one-way transmission of linearly polarized waves in such systems. Furthermore, an optical isolation of circularly polarized waves was investigated in both nonlinear chiral metastructures~\cite{Shadrivov} and in a thin magneto-photonic crystal slab~\cite{Fan_isolat}. However, both systems have limitations in their applicability; for example, the nonlinear chiral metastructure operates at microwave frequencies~\cite{Shadrivov}, and the extension of this interesting idea to higher frequencies would be challenging due to the weak optical nonlinearities. On the other hand, the elegant idea of optical isolation in perforated magneto-optical slabs~\cite{Fan_isolat} might experience difficulties with fabrication due to the mechanical stability of magneto-optical materials.

Merging the concepts of metamaterial design with magneto-optical response might open possibilities for engineering novel nonreciprocal systems for optical isolation and signal processing. Metamaterials -- artificially engineered heterostructures with elements much smaller than operating wavelengths~\cite{Pendry,Soukoulis} -- can be treated macroscopically as a homogeneous media exhibiting exotic values for permittivity and permeability, such as negative refractive index~\cite{Smith,Shalaev,Valentine} naturally not observed in nature. Recently, materials with epsilon-near-zero (ENZ) permittivity were proposed and fabricated~\cite{Silverinha,Alu_ENZ,Brian_ENZ}. One of the most intriguing properties of these materials is the nearly zero phase progression of propagating waves, which can be employed for the efficient energy tunneling~\cite{Silverinha,Brian_ENZ} and for the design of the next generation of waveguides~\cite{Brian_circ}. What is more important in these materials is that a small variation of their parameters may allow tuning their properties form being opaque ``metals" (with $\textrm{Re}(\varepsilon)<0$) to transparent dielectrics (when $\textrm{Re}(\varepsilon)>0$).

In this Letter, we suggest a design approach for optical isolation for circularly polarized waves by utilizing both the epsilon-near-zero materials or structures and the magneto-optical materials. We show theoretically that magnetically active ENZ media is transparent for one of the circularly polarized waves, whereas it is opaque for another polarization handedness. We propose two realistic examples where such a scenario is feasible. Specifically, we study the case of a stack of alternating metal-dielectric slabs, and the case of a waveguide filled with magneto-optical media operating near its cut-off frequency. We demonstrate numerically that the isolation of circularly polarized states is possible in both of these example structures.

The paper is organized as follows. In Section II we give a generalized concept for designing the ENZ-MO isolator for circularly polarized waves. We discuss two particular examples of such isolator in Section III. In particular, we consider a metal-dielectric stack and a waveguide near its cut-off, and show that both structures exhibit effective ENZ properties combined with gyrotropy allowing isolation of one of the circularly polarized states when magneto-optical activity is present. Section IV provides the conclusion.

\section{Heuristic approach for the ENZ isolator deign}

\begin{figure}[h]
\begin{center}
\includegraphics[width=\columnwidth]{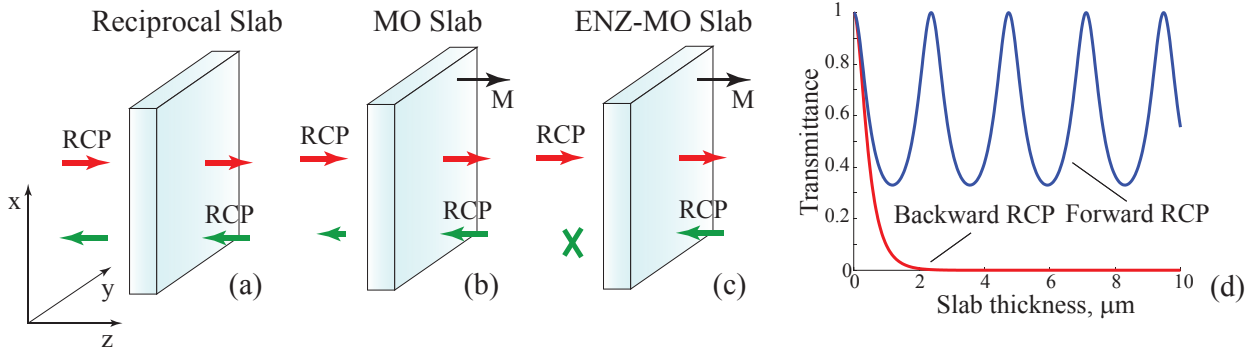}
\end{center}
\caption{Schematic of RCP forward and backward transmission through a) reciprocal, b) magneto-optical, and c) magnetically active epsilon-near-zero slabs, respectively. d) Transmittance of the forward and backward RCP plane waves through ENZ-MO slab as a function of the slab thickness.} \label{slabs}
\end{figure}

We begin our analysis with a brief insight into the properties of a conventional magneto-optical materials. Consider, for example, a slab of magneto-optical medium with a magnetization direction normal to the slab interface, see Fig.~\ref{slabs}(b). In this case the MO relative permittivity can be described by an antisymmetric tensor~\cite{Landau}:

 \begin{equation}\label{epsilon}
  \bar{\bar{\varepsilon}}= \left( \begin{array}{ccc}
   \varepsilon_{mo} & -i\alpha & 0 \\
   i\alpha & \varepsilon_{mo} & 0 \\
   0 & 0 & \varepsilon_\bot \end{array} \right),
\end{equation}
where $\varepsilon_{mo}$ and $\varepsilon_\bot$ are diagonal components of relative permittivity tensor, $\alpha>0$ is the off-diagonal component of the permittivity tensor responsible for the ``strength'' of magneto-optical activity of the medium. At optical frequencies, the parameter $\alpha$ is often very small and usually of the order of $10^{-2}$ or even smaller.

For plane waves propagating along the magnetization direction (i.e. in Faraday configuration) the eigenmodes in magneto-optical media are two circularly polarized waves: $E_x\pm iE_y$. Solving Maxwell's equations for the circularly polarized waves propagating in the magneto-optical media in forward direction we get a well known relation for the corresponding wavevectors~\cite{Landau}: $k_{LCP}^+ = (\omega/c) \sqrt{\varepsilon_{mo}-\alpha}$ for left-handed circular polarized waves and $k_{RCP}^+ = (\omega/c) \sqrt{\varepsilon_{mo}+\alpha}$ for right-handed circular polarized waves. To account for the backward propagating waves the time reversal operation ($t \rightarrow -t$) needs to be performed. In this case the change of sign in front of $\alpha$ has to be accounted for in the wave dispersions:  $k_{LCP}^- = (\omega/c) \sqrt{\varepsilon_{mo}+\alpha}$ and $k_{RCP}^- = (\omega/c) \sqrt{\varepsilon_{mo}-\alpha}$ for backward LCP and RCP waves, respectively. Clearly, backward and forward propagating circular polarized waves of the same handedness have different wave numbers with differing dispersion ($|k^+|\neq |k^-|$) yielding difference in their transmittance and reflectance~\cite{Landau}.

Exploiting the metamaterial concept, one can engineer metamaterials and structures in order to deliberately tailor the value of permittivity function~\cite{Soukoulis}. In Refs.~\cite{Brian_ENZ,Brian_circ,Salandrino} it was shown that in properly designed structures, one can achieve $\varepsilon\rightarrow0$, and consequently very small phase progression ($k\rightarrow 0$). Examples of such structures include waveguides near their cut-off frequency for the TE modes~\cite{Brian_ENZ}, layered metal-dielectric stacks~\cite{Salandrino} and natural materials at frequencies for which the real part of their permittivity is near zero (while its imaginary par is small)~\cite{Johnson}. By integrating such epsilon-near zero structures with magneto-optical materials it may be possible to design effective magneto-active media with nulled diagonal components in the permittivity tensor, i.e. $\varepsilon_{mo}\rightarrow0$ in Eq.(\ref{epsilon}). Assuming now that such a case is possible, we obtain following dispersion relations for circularly polarized wave eigenmodes:

\begin{gather}
   \textrm{forward} \;\;\;\;\;\;\;\;\;\;\;\; k_{LCP}^+ = (\omega/c) \sqrt{-\alpha} \;\;\;\;\;\;\;\;\;\; \textrm{and} \;\;\;\;\;\;\;\;\; k_{RCP}^+ = (\omega/c) \sqrt{\alpha} \nonumber \\
   \textrm{backward} \;\;\;\;\;\;\;\;\;\;\;\; k_{LCP}^- = -(\omega/c) \sqrt{\alpha} \;\;\;\;\;\;\;\;\;\; \textrm{and} \;\;\;\;\;\;\;\;\; k_{RCP}^- = (\omega/c) \sqrt{-\alpha}.
\end{gather}

Assuming a lossless condition, as evident from Eq.(2), for $\alpha>0$ the forward propagating LCP waves and backward propagating RCP waves have imaginary wavevectors in such a medium, implying that the medium is opaque for these waves, while for the two other eigenmodes, the wave numbers are real, and thus the medium is transparent to these waves. Consequently, it would be possible to achieve two effects simultaneously: 1) isolate forward and backward propagating waves with the same handedness (see schematic on Fig.~\ref{slabs}(c)), and 2) isolate LCP and RCP waves propagating in the same direction.

In Fig.~\ref{slabs}(d) we plot the transmittance of the RCP wave in forward and backward directions through a slab made of such material with the variation of slab thickness. To be more specific we consider here the operating wavelength to be $\lambda=1.5\mu$m, ambient medium is air and $\alpha=0.1$. Note that, $\alpha$ is usually smaller than this value, for example, for bismuth iron garnet, a commonly used magneto-optical material, $\alpha=0.06$ at $\lambda=1.55\mu$m. However, here we intentionally consider slightly higher values of magneto-optical activity in order to stress and highlight the observed effects. We see that the transmission of backward RCP wave exponentially decays as the slab thickness increases, and for the $2\mu$m thick slab it already falls below $1\%$. At the same time the slab acts as a Fabry-Perot resonator for the forward propagating RCP with a first transmission resonance at $\sim2.3\mu$m thick slab. Analogously the forward propagating LCP would be suppressed similarly to backward RCP. Hence, engineering a magnetically active ENZ slab would allow to achieve isolation of circularly polarized waves with a slab thickness comparable with the incident wavelength. In the following section we propose to possible realizations of such magneto-active ENZ isolators.

\section{Design of ENZ isolator}

In order to engineer the magneto-active ENZ media we follow the two methods developed in the area of ENZ metamaterials. In the first approach we consider a metal-dielectric stack~\cite{Salandrino}, shown schematically in Fig.~\ref{structure}(a). For the second method, a rectangular waveguide operating at the cutoff frequency of its TE mode is utilized~\cite{Brian_ENZ}, see Fig.~\ref{structure}(b).

\begin{figure}[h]
\begin{center}
\includegraphics[width=0.7\columnwidth]{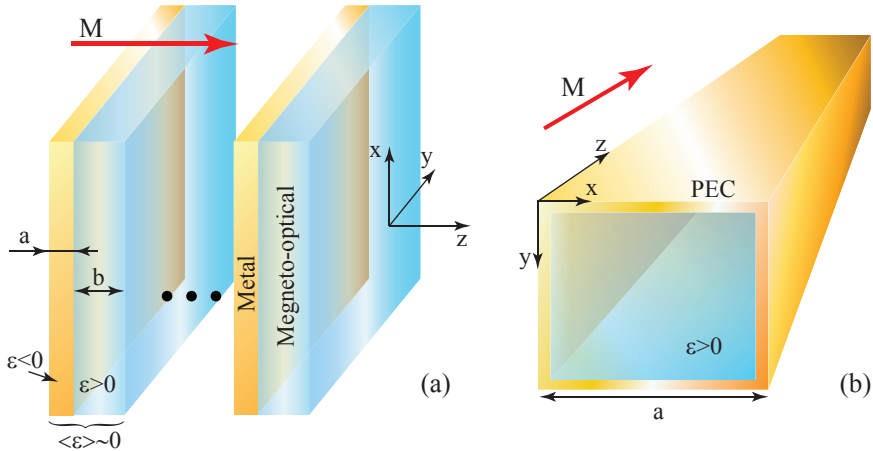}
\end{center}
\caption{Schematic of the magneto-active structures exhibiting effective epsilon-near-zero permittivity: a) periodic metal-MO stack, and b) rectangular metallic (PEC) waveguide with a square cross-section. Red arrow shows the direction of magnetization.} \label{structure}
\end{figure}

\subsection*{Metal-dielectric stack}

We consider a stack of periodically alternating metal and magneto-optical slabs with thicknesses $a$ and $b$, respectively (Fig.~\ref{structure}(a)). When the thickness of the slabs is much smaller than the wavelength of operation the effective-medium approach may be used~\cite{Salandrino}. Using the Cartesian coordinate systems shown in Fig.~\ref{structure}, the $x$ and $y$ components of the effective medium permittivity for an unbiased system, i.e. when $\alpha = 0$, are given by: $\varepsilon_{xx}=\varepsilon_{yy} = (a\varepsilon_{met}+b\varepsilon_{MO})/(a+b)$. By properly choosing these parameters, we can achieve near-zero permittivity for $\varepsilon_{xx}$ and $\varepsilon_yy$. Then according to the predictions made in Section II we expect that such a stack, when magnetically biased, behaves as an isolator for circularly polarized waves. To prove our predictions we perform full-wave numerical simulations with the CST Microwave Studio$^{TM}$~\cite{CST}.

For the purpose of numerical demonstration here we consider the magneto-optical layers to be made of bismuth iron garnet with $\varepsilon_{MO}=6.25$ and $\alpha=0.1$, and the metal to be lossless with relative permittivity $\varepsilon=-12.5$. In this case choosing $a = (1/2) b$ we obtain zero effective epsilon for $\varepsilon_{xx}$ and $\varepsilon_{yy}$. A plane wave with free space wavelength of $1.5 \mu$m is incident normally on a stack containing $64$ unit cells. We choose the thickness of the metal layer to be $a=20$nm and that of magneto-optical slab to be $b=40$nm, so that the overall stack thickness of the entire structure is $3.84 \mu$m. Such choice of slab thicknesses allows the applicability of the effective medium approximation, i.e. $a+b \ll \lambda$.

\begin{figure}[h]
\begin{center}
\includegraphics[width=\columnwidth]{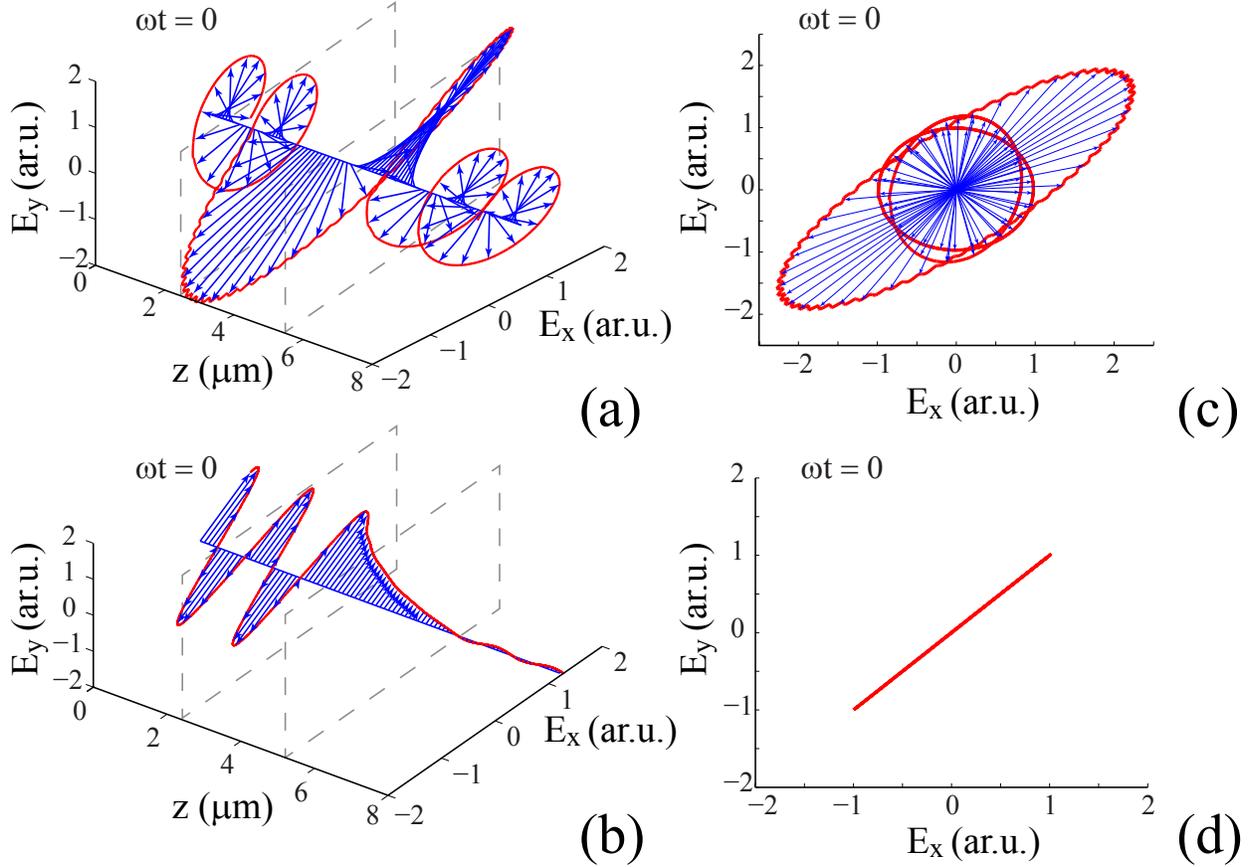}
\end{center}
\caption{a) and b) instantaneous electric field distribution for forward propagating RCP and LCP, respectively (Media 1). c) and d) front views of the electric field for forward RCP and LCP, i.e. projections of the fields depicted in panels (a) and (b) on the ($x-y$) plane, correspondingly.}\label{stack}
\end{figure}

Figs.~\ref{stack}(a,b) show instantaneous distributions of the electric field vector for the RCP and LCP forward propagating plane waves at the specific time instant $(\omega t=0)$. We observe that RCP wave propagates through the stack and preserves its amplitude, Fig.~\ref{stack}(a), whereas LCP is almost completely reflected and only small fraction of it is tunneled trough the stack, Fig.~\ref{stack}(b). Note that, the incident LCP and its reflection from the stack form a standing circularly polarized wave. Instantaneously it looks like linearly polarized wave, however the polarization plane rotates with time (not shown here). To analyze the the degree of polarization we plot the front view of the electric field vector, i.e. its projection on the plane transverse to the direction of propagation ($x-y$ plane) at the same specific time instant $(\omega t=0)$, Fig.~\ref{stack}(c,d). Clearly, RCP preserves its polarization handedness with propagation, i.e. stays circular after the transmission through the stack, Fig.~\ref{stack}(c). At the same time reflected LCP, as was noted, forms an instantaneous linearly polarized state (which is rotating with time), Fig.~\ref{stack}(d). We note that the backward propagating RCP behaves similarly to the forward LCP, i.e., in particular it is reflected from the stack, indicating that the stack acts as an isolator for circularly polarized waves.

\subsection*{Waveguide at cut-off frequency}

It is known that a metallic rectangular waveguide with the perfectly conducting walls, near its cut-off frequency for TE10 mode effectively behaves as an ENZ strcture~\cite{Brian_ENZ}. We show that such a waveguide when filled with a magneto-optical material  may also also act as an ENZ isolator.

We consider a rectangular waveguide with a square cross-section ($10\mu$m$\times10\mu$m) filled with bismuth iron garnet ($\varepsilon_{mo}=6.25$ and $\alpha=0.1$), see Fig.~\ref{structure}(b). Note that the direction of magnetization in this case is along the waveguide, i.e. along the direction of mode propagation. The square cross section of the waveguide leads to the degeneracy of the lower order TE modes ($TE_{01}$ and $TE_{10}$). For the given waveguide dimensions both modes exhibit cut-off at $6$THz (corresponding wavelength of $50 \mu$m).

\begin{figure}[h]
\begin{center}
\includegraphics[width=\columnwidth]{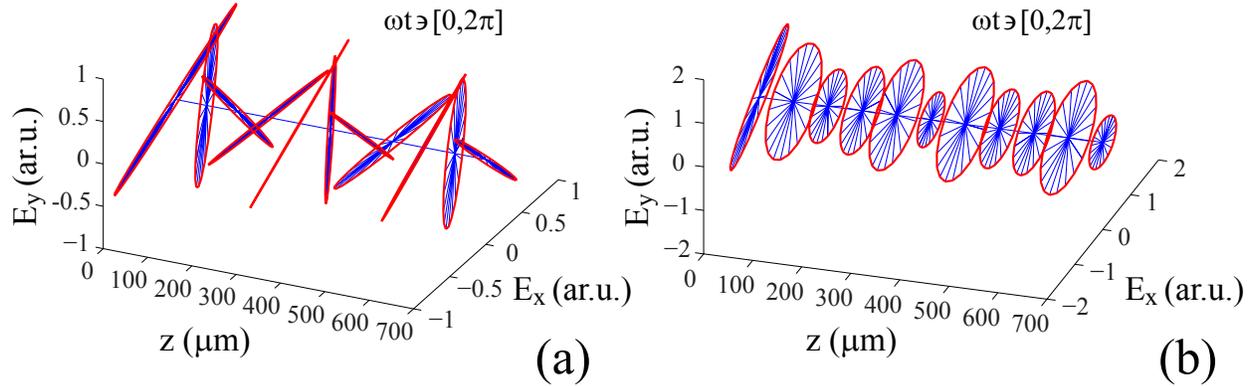}
\end{center}
\caption{Electric field distribution along the waveguide at different cross sections and different time frames (Media 2): a) above cut-off at $6.1$THz and b) below cut-off at $5.98$THz.} \label{waveguide}
\end{figure}

When magneto-optical activity is introduced these degenerate states couple and form two new conjugate eigen-states $\psi_1 \pm i\psi_2$. Note that such coupling is possible only between orthogonal states existing simultaneously in the unbiased waveguide, for example, a single mode rectangular waveguide does not allow any rotating states. These new conjugate modes can be considered as analogues of LCP and RCP plane waves as the eigen-modes of the bulk media. Using this analogy we define these states as left-rotating and right-rotating eigen-modes respectively. The dispersion relations defining the eigenvalues of right and left rotating states are given as: $k_{\pm} = \sqrt{(\omega^2/c^2)(\varepsilon\pm\alpha)-\pi^2/a^2}$, where $a$ is the waveguide's edge dimension. Hence near the cut-off frequency, when $(\omega/c)\sqrt{\varepsilon}=\pi/a$, we arrive to the conditions similar to that discussed in Section II. In particular, we see that near the cut-off frequency, one of the rotating states is not propagating, whereas the same rotating state propagates in the reverse direction.

To demonstrate this property of electromagnetic wave isolation inside the rectangular waveguide filled with the MO material, we perform a series of full-wave numerical calculations. In our simulations we assume that the waveguide walls are made of perfect electric conductor (PEC). We excite the waveguide with $TE_{10}$ mode and study the corresponding field distribution and transmission characteristics. For magneto-optically active case the $TE_{10}$ is decomposed into the counter-rotating eigen-modes supported by magnetized waveguide. Above the cut-off frequency both counter-rotating states exist, and due to the difference between their wavenumbers the phase between the $TE_{10}$ and $TE_{01}$ is accumulated with propagation along the waveguide, similarly to the Faraday effect in bulk media~\cite{Landau}. Below cut-off only one of the counter-rotating states exists, hence $TE_{10}$ launched into the waveguide would generate this rotating state.

We trace the time variation of electric field at several different positions along the waveguide, i.e. along the central line, see Figs.~\ref{waveguide}(a,b). Fig.~\ref{waveguide}(a) shows the electric field variation above cut-off at $6.1$ THz. We observe that in this case the polarization is highly elliptical, almost linear. Furthermore, we notice rotation of the polarization axis as we move along the waveguide as predicted earlier. Below the cut-off at $5.98$THz, see Fig.~\ref{waveguide}(b), we observe propagation of purely circularly polarized state, which clearly manifests the isolation of one of the rotating states. Finally, we note that the small deviation from the linear polarization on Fig.~\ref{waveguide}(a) and modulation of the amplitude on Fig.~\ref{waveguide}(b) is due to small reflection appearing due to the numerical mismatch at the input and output ports of the waveguide.

\section{Conclusion and acknowledgements} We showed that magnetically active epsilon-near-zero structures acts as compact isolator for circularly polarized waves. We have demonstrated two cases for such media: a metal-dielectric stack and a rectangular waveguide near the cut-off frequency. Using full-wave numerical simulation, we have shown that in both structures the complete suppression of one of the circularly polarized states is possible.

This work is supported in part by the US Air Force Office of Scientific Research (AFOSR) grant number FA9550-10-1-0408.

\end{sloppy}
\end{document}